\documentclass[noinfoline,stslayout]{imsart}

\usepackage{amsfonts,amssymb,amsmath,amscd,amsthm,latexsym}
\usepackage{natbib}
\usepackage{graphicx}
\usepackage[]{units}
\usepackage{algorithmicx}
\usepackage{algpseudocode}
\usepackage{qtree}
\usepackage{algorithm}
\usepackage{xcolor}

\begin{document}

\begin{frontmatter}
\small
\title{A discussion of ``Bayesian model selection based on proper scoring rules" by A.P. Dawid and M.
Musio\protect\thanksref{T1}}
\author{C. Grazian${}^{1,2}$, I. Masiani${}^{1,2}$, and C.P. Robert${}^{2,3}$}
\affiliation{${}^1$Universit\`a La Sapienza--Roma, ${}^2$Universit\'e Paris-Dauphine, and ${}^3$University of Warwick}
\begin{abstract}
This note is a discussion of the article ``Bayesian model selection based on proper scoring rules" by A.P. Dawid and M.
Musio, to appear in {\em Bayesian Analysis}. While appreciating the concepts behind the use of proper scoring rules,
including the inclusion of improper priors, we point out here some possible practical difficulties with the advocated approach.
\end{abstract}

\thankstext{T1}{This work is based on a Master research project written by the second author under the joint supervision of the
first and third authors, at Universit\'e Paris-Dauphine. The first author is a PhD candidate at Universit\`a La
Sapienza--Roma and Universit\'e Paris-Dauphine.}
\end{frontmatter}

The frustrating issue of Bayesian model selection preventing improper priors \citep{degroot:1982} and hence most objective Bayes
approaches has been a major impediment to the development of Bayesian statistics in practice \citep[see, e.g.,][]{marin:robert:2007}, as the
failure to provide a ``reference" answer is an easy entry for critics who point out the strong dependence of posterior
probabilities on prior assumptions. This was presumably not forecasted by the originator of the Bayes factor, Harold
Jeffreys, who customarily used improper priors on nuisance parameters in his construction of Bayes factors
\citep{robert:chopin:rousseau:2009} (The expansion (4) in the paper, while worth recalling, is unlikely to convince such
critics.) While others may object to the use of improper and non-informative priors in such settings \citep{kadane:2011},
we consider it a very welcome item of news that a truly Bayesian approach can allow for improper priors. 

As also pointed out in the paper, there exists a wide range of ``objective Bayes" solutions in the literature
\citep[see, e.g.,][]{robert:2001}, all provided with validating arguments of sorts, but this range by itself implies that such
solutions are doomed in that they cannot agree for a given dataset and a given prior. 

Finding a criterion that does not depend on the normalising constant of the predictive possibly is the unravelling key to handle
improper priors and we congratulate the authors for this finding of the Hyv\"arinen score and related proper scoring
rules. Some difficulties deriving from the use of improper prior distributions in model choice may be solved by applying
the approach proposed in the paper. There are nonetheless some issues with this solution:
\begin{itemize}

\item calibration difficulties: once the score value is computed for a collection of models, the calibration of their 
respective strengths very loosely relates to a loss function, hence the approach makes informed decision in favour of a 
model difficult or prone to subkective choices;

\item a clear dependence on parameterisation: changing the observation $x$ into the bijective
tranform $\mathfrak{h}(x)$ produces a different score;

\item a similar dependence on the dominating measure: as exhibited in the case of exponential families and eqn.~(30), changing
the dominating measure into an equivalent dominating measure modifies the score function;

\item the arbitrariness of the Hyv\"arinen score itself, which is indeed independent of the normalising constant. The
article offers a limited collection of arguments in favour of this particular combination of derivatives. Since there exist a immense range of
possible score functions, a stronger connection with inferential properties is a clear requirement;

\item as noted above, consistency is not a highly compelling argument for the layperson, as it does not help in the
calibration or selection of the score function. Exhibiting the multivariate score as being inconsistent, while the prequential score
remains consistent, is highlighting this difficulty and may deter users from adopting the approach.
\end{itemize}

Furthermore, the only application of the method presented in the paper is within the setting of the Normal
linear model and we worry that the approach may not be easily extended to other types of models. In particular, the representation
of the precision matrix of the marginal distribution in equation (33), based on the Woodbury matrix inversion lemma, is
essential to easily apply the proper scoring rule approach to model choice with an improper prior, given that an improper
prior may then be seen as a limiting version of a conjugate prior and its influence disappears in the following computations.
However, the approach overcomes the singularity of the precision matrix of the marginal distribution. 

We performed several simulation studies when applying the method to models that differ from the Normal linear model.
When chosing between two different models with no covariates, we observed that the proposed approach may perform well as
for instance when a Gamma model is compared with a Normal model (when comparing with a Bayes factor). However, when the
comparison is operated between a Pareto distribution and a Normal distribution, the approach does not customarily reach 
the right model when data are generated from the Pareto distribution, while the Bayes factor always leads the right model. 
In addition, the method based on the Hyv\"arinen scoring rule may not be applied with some models, for example when data 
come from a Laplace distribution, which is not differentiable in $0$, or for discrete models. 

\begin{figure}[tb]
\centering
\includegraphics[width=1.1\textwidth]{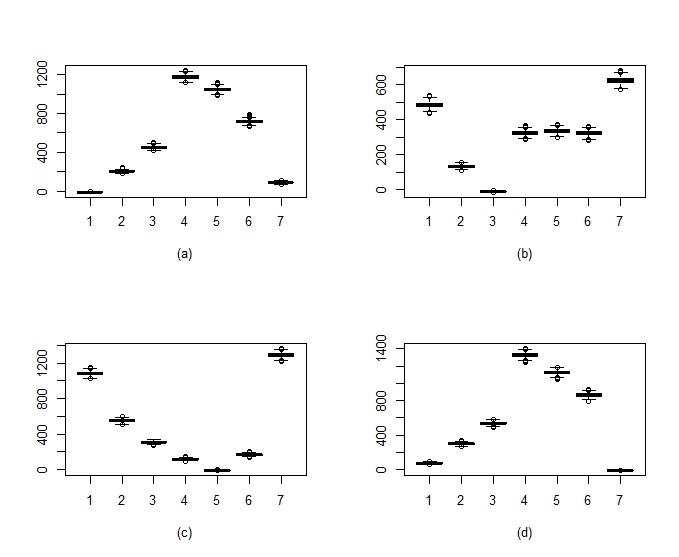}
\caption[Nested models: boxplots of the scores]{{\it Boxplots over 1,000 simulations of the sample distributions of the scores of seven models under analysis, varying the true model. Model selection is performed in the case of nested linear normal models, where the data ${\bf y}=(y_1,...,y_n)$ have distribution ${\bf y}|\boldsymbol\theta\sim N({\bf X}\boldsymbol\theta, \sigma^2)$, where $\bf X$ is the design matrix, $n=100$ and $\sigma^2=10$. In (a) the true model is $M_1:= {\boldsymbol\theta}_1=(1, 0 , 0 , 0 , 0 , 0)$, in (b) is $M_3 := {\boldsymbol\theta}_3=(1,1,1,0,0,0)$, in (c) is $M_5 := {\boldsymbol\theta}_5=(1,1,1,1,1,0)$ while in (d) the correct model is $M_7 := {\boldsymbol\theta}_7=(0,0,0,0,0,0)$.}} 
\label{bxp}
\end{figure}

\begin{figure}[tb]
\includegraphics[width=1.0\textwidth]{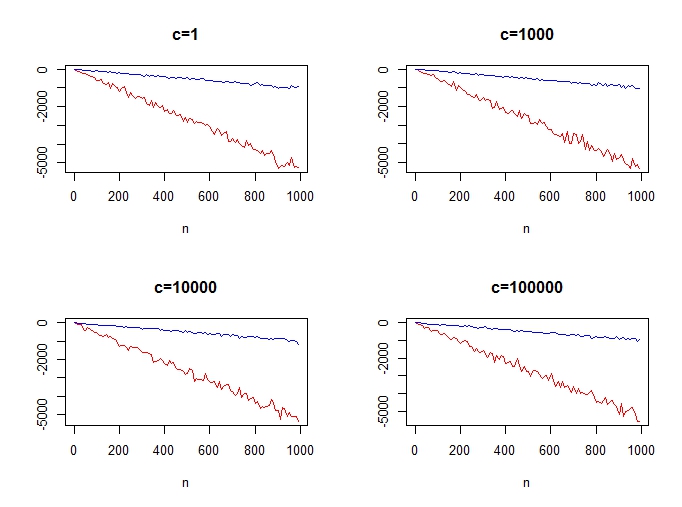}
\caption[Linear model: comparison between the Bayes factor approach and the one based on the score functions when
increasing the sample size and the prior variance]{{\it Linear model: log-Bayes factor (red) and difference of the
scores (blue) in function of the increasing sample size $n=1,...,1000$, and of the prior variance on $\theta$,
$V=c\sigma^2$, where $\sigma^2=10$ and known. Given data ${\bf y}=(y_1,...,y_n)$ with distribution ${\bf y}|\theta\sim
N({\bf X}\theta, \sigma^2)$ we consider two possible models generating the data: $M_0 :\, \theta=(0, 1)$ and $M_1 :\,
\theta=(1, 0)$ when the true model is $M_1$.}}
\label{res1c}
\end{figure}

\begin{figure}[tb]
\includegraphics[width=1.0\textwidth]{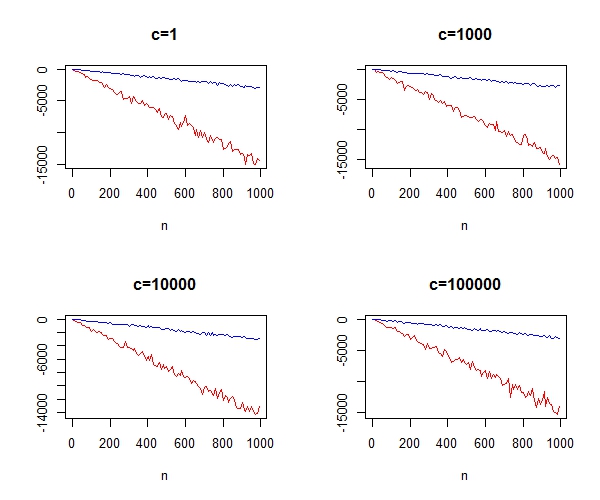}
\caption[Nested models: comparison between the Bayes factor approach and the one based on the score functions when
increasing the sample size and the prior variance]{{\it Nested models: log-Bayes factor (red) and difference of the
scores (blue) in function of the increasing sample size $n=1,...,1000$, and of the prior variance on $\theta$,
$V=c\sigma^2$, where $\sigma^2=10$. The setting is the same as Figure \ref{bxp} and we have to choose between model $M_3
:\, {\boldsymbol\theta}=(1,1,1,0,0,0)$ and model $M_6 :\, {\boldsymbol\theta}=(1,1,1,1,1,1)$, $M_6$ being the true
model.}}
\label{resnc}
\end{figure}

Our simulation studies have also dealt with linear models, nested and not nested. The performance of the multivariate
score when comparing Normal linear models is excellent, as shown in Figure \ref{bxp}, even using an improper prior,
provided the sample size is larger than the number of parameters in the model: following repeated simulations, we
observed that the method is always able to choose the right model. Although this approach shows a consistent behavior
and chooses the right model with higher and higher certainty when the sample size increases, our simulations have also
shown that the log proper scoring rule tends to infinity more slowly than the Bayes factor or than the likelihood ratio.
It is approximately four times slower, all priors being equal, as shown in Figures \ref{res1c} and \ref{resnc}, which
represent the comparison between the approach based on the log-Bayes factor and the one based on the difference between
the score functions for the case of linear model, both nested (Fig. \ref{resnc}) and non-nested (Fig. \ref{res1c}).

%\begin{align*}
%M_1 &:= {\boldsymbol\theta}_1=(1, 0 , 0 , 0 , 0 , 0) \\
%M_2 &:= {\boldsymbol\theta}_2=(1,1,0,0,0,0) \\
%M_3 &:= {\boldsymbol\theta}_3=(1,1,1,0,0,0) \\
%M_4 &:= {\boldsymbol\theta}_4=(1,1,1,1,0,0) \\
%M_5 &:= {\boldsymbol\theta}_5=(1,1,1,1,1,0) \\
%M_6 &:= {\boldsymbol\theta}_6=(1,1,1,1,1,1) \\
%M_7 &:= {\boldsymbol\theta}_7=(0,0,0,0,0,0)
%\end{align*}

As a final remark, we would like to point out the alternative proposal of \cite{kamary:mengersen:x:rousseau:2014} for
correctly handling partly improper priors in testing settings through the tool of mixture modelling, each model under
comparison providing a component of the mixture. Therein, the authors show consistency in a wide range of situations. We
see the approach through mixtures as more compelling for the many reasons provided in the paper, in particular as the
posterior distribution of the weight of a model is easily interpretable and scalable towards selecting this model or its
alternative. Allowing improper priors solely on the nuisance parameters, that is, on the parameters not being tested
sounds to us like another convincing feature of the mixture approach.
 
%\bibliographystyle{ims}
%\bibliography{biblio}
\hyphenation{Post-Script Sprin-ger}

\end{document}